\documentclass[prl,longbibliography,showpacs,twocolumn,superscriptaddress,amsmath,amssymb,verbatim]{revtex4-1}
\usepackage{graphicx}
\usepackage{lmodern}
\usepackage{epstopdf}
\usepackage{dcolumn}
\usepackage{bm}
\usepackage{subfigure}
\usepackage{textcomp}
\usepackage[pdftex,colorlinks=true,citecolor=blue,linkcolor=blue,urlcolor=blue,bookmarks=true]{hyperref}
\usepackage[sort&compress]{natbib}

\newcommand{\Nd}{NdTa$_7$O$_{19}$}
\begin{document}
\preprint{APS}

\title{Quantum spin liquid in the Ising triangular-lattice antiferromagnet\\ neodymium heptatantalate}

\author{T. Arh}
\affiliation{Jo\v{z}ef Stefan Institute, Jamova c. 39, 1000 Ljubljana, Slovenia}
\affiliation{Faculty of Mathematics and Physics, University of Ljubljana, Jadranska u. 19, 1000 Ljubljana, Slovenia}
\author{B. Sana}
\affiliation{Department of Physics, Indian Institute of Technology Madras, Chennai 600 036, India}
\author{M. Pregelj}
\affiliation{Jo\v{z}ef Stefan Institute, Jamova c. 39, 1000 Ljubljana, Slovenia}
\author{P. Khuntia}
\affiliation{Department of Physics, Indian Institute of Technology Madras, Chennai 600 036, India}
\author{Z. Jagli\v{c}i\'c}
\affiliation{Faculty of Civil and Geodetic Engineering, University of Ljubljana, 1000 Ljubljana, Slovenia}
\affiliation{Institute of Mathematics, Physics and Mechanics, 1000 Ljubljana, Slovenia}
\author{M. D. Le}
\affiliation{ISIS facility, Rutherford Appleton Laboratory, Chilton, Didcot, OX11 0QX, Oxfordshire, UK}
\author{P. Biswas}
\affiliation{ISIS facility, Rutherford Appleton Laboratory, Chilton, Didcot, OX11 0QX, Oxfordshire, UK}
\author{A. Ozarowski}
\affiliation{National High Magnetic Field Laboratory, Florida State University, Tallahassee, Florida 32310, USA}
\author{A. Zorko}
\email{andrej.zorko@ijs.si}
\affiliation{Jo\v{z}ef Stefan Institute, Jamova c. 39, 1000 Ljubljana, Slovenia}
\affiliation{Faculty of Mathematics and Physics, University of Ljubljana, Jadranska u. 19, 1000 Ljubljana, Slovenia}


\begin{abstract}

{\bf
Disordered magnetic states known as spin liquids are of paramount importance in both fundamental and applied science.
A classical state of this kind was predicted for the Ising antiferromagnetic triangular model more than half a century ago, while additional non-commuting exchange terms were proposed to induce its quantum version -- a quantum spin liquid. 
These predictions have not been yet confirmed experimentally. 
Here we report a discovery of such a state in the structurally perfect triangular-lattice antiferromagnet NdTa$_7$O$_{19}$.
Its magnetic ground state is characterized by spin-1/2 degrees of freedom with Ising-like interactions and gives rise to low-frequency spin excitations persisting down to the lowest temperatures.
Our study demonstrates the key role of strong spin-orbit coupling in stabilizing spin liquids that result from magnetic anisotropy and highlights the large family of rare-earth (RE) heptatantalates RETa$_7$O$_{19}$ as a novel framework for realization of these states, which represent a promising platform for quantum applications. 
}

\end{abstract}

\maketitle

Quantum materials that are characterized by emergent quantum phenomena leading to novel functionalities represent a mainstream in modern condensed matter research \cite{basov2017towards,keimer2017physics,tokura2017emergent,wen2019choreographed}. 
One such intriguing phenomenon is a quantum entangled but magnetically disordered state known as quantum spin liquid (SL), which is a natural ground-state contender on geometrically frustrated spin lattices \cite{balents2010spin, savary2017quantum, zhou2017quantum, broholm2020quantum}. 
Initially, it was proposed for a triangular-lattice antiferromagnet (TAFM) with Heisenberg, i.e., isotropic nearest-neighbour exchange interactions \cite{anderson1973resonating}, but theory later converged on a magnetically ordered ground state \cite{capriotti1999long}.
Nevertheless, deviations from this model, such as next-neighbour \cite{iqbal2016spin,hu2019dirac} and spatially anisotropic exchange interactions \cite{yunoki2006two,heidarian2009spin}, or magnetic anisotropy \cite{yamamoto2014quantum,zhu2018topography, maksimov2019anisotropic} can still stabilize spin liquids.
In fact, the most anisotropic, i.e., the classical Ising TAFM model including only the out-of-plane component of the exchange interaction ${\cal J}_z$, was the first spin model predicted to possess a disordered, macroscopically degenerated ground state -- a classical SL -- by Wannier already in 1950 \cite{wannier1950antiferromagnetism}.
An additional in-plane component ${\cal J}_{xy}$ introduces quantum fluctuations \cite{moessner2001ising} that lift the degeneracy of the ground-state manifold and eventually select
magnetic order \cite{yamamoto2014quantum}. 
However, close to the Ising limit a quantum SL can potentially be realized \cite{diep2004frustrated, fazekas1974ground}. 
Neither the classically disordered Wannier state nor its disordered quantum successor characterized by quantum fluctuations have been so far confirmed experimentally, which thus remains one of the fundamental quests in the field.    

Materials with strongly spin-orbit entangled local magnetic moments represent a natural habitat for SL states borne out of magnetic anisotropy on frustrated spin lattices \cite{li2016anisotropic, iaconis2018spin}, as well as for various other intriguing many-body quantum phenomena \cite{witczak2014correlated}.
The rare-earth (RE) based YbMgGaO$_4$ compound is the most intensively investigated TAFM example of this kind \cite{li2015rare, shen2016evidence, paddison2017continuous, li2016muon}.
However, its SL-like ground state likely emerges from structural disorder \cite{zhu2017disorder, kimchi2018valence}, similarly as proposed also for other well studied TAFM materials lacking magnetic order \cite{watanabe2014quantum, kawamura2019nature}, including organic charge-transfer salts \cite{shimizu2003spin, itou2008quantum} and 1T-TaS$_2$ \cite{klanjsek2017high}.
As the origin of the apparent SL character in all these materials remains debatable, new disorder-free TAFM candidates are highly desired, especially RE-based ones \cite{liu2018rare, bordelon2019field, ashtar2019reznal}, where large magnetic anisotropy could warrant the existence of SL ground states.
Here we report the realization of such a state in a newly discovered RE-based TAFM material \Nd~with perfect triangular symmetry.
Its crystal-electric-field (CEF) ground-state Kramers' doublet is characterized by highly anisotropic exchange interactions between the effective spin-1/2 degrees of freedom, ${\cal J}_z=0.90(2)$\,K and ${\cal J}_{xy}=0.16(2)$\,K.
In absence of any structural disorder, these Ising-like interactions are likely the key for stabilizing its quantum-disordered ground state with characteristic low-frequency excitations that are found to persist down to at least 66\,mK.
\begin{figure*}[!]
\centering
\includegraphics[width=0.75\linewidth]{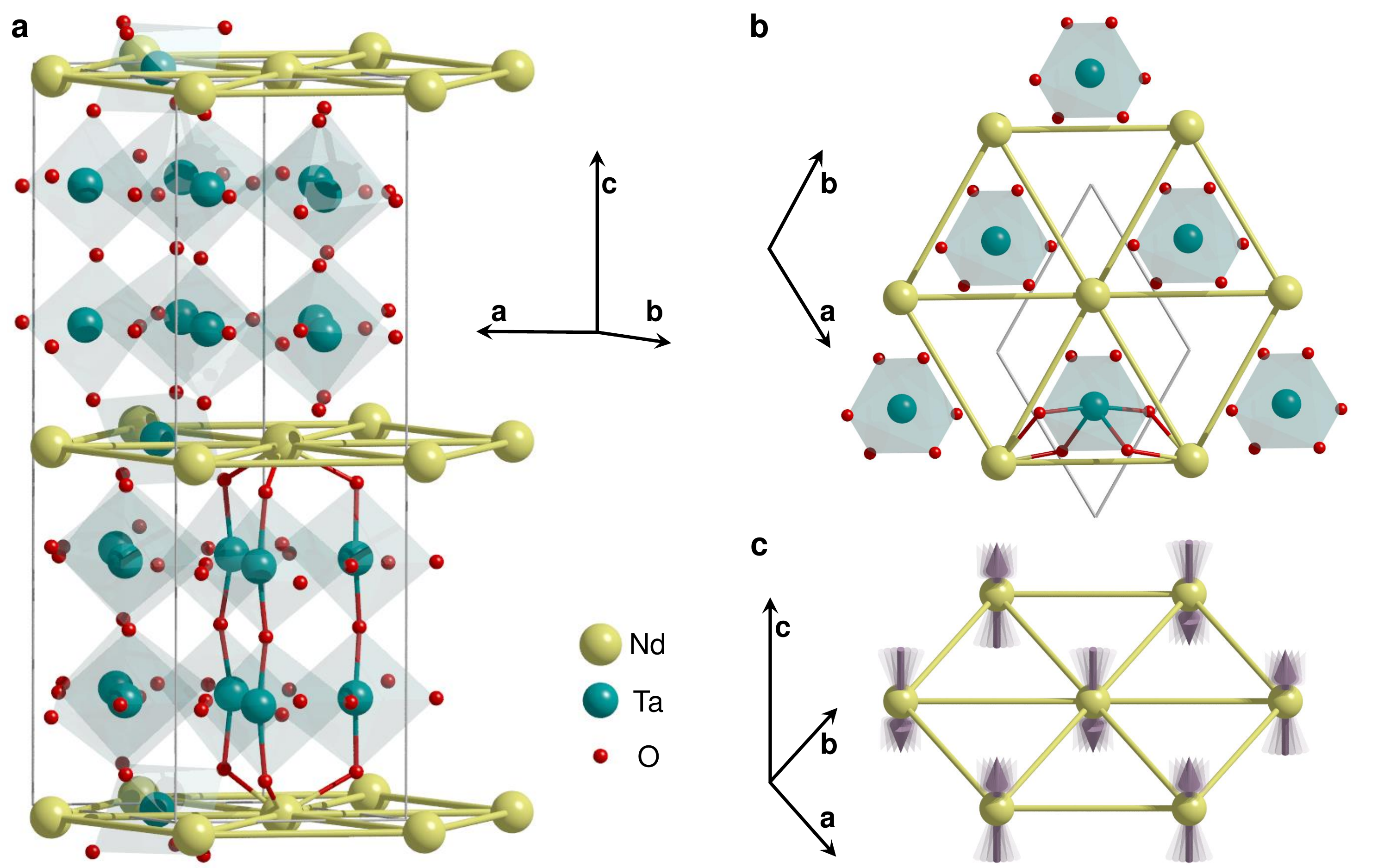}
\caption{{\bf Crystal structure and magnetic ground state of NdTa$_7$O$_{19}$:} {\bf a,b}, In the hexagonal crystal structure of NdTa$_7$O$_{19}$ (space group $P\bar{6}c2$, thin gray lines show the unit cell) the two-dimensional triangular layers of magnetic Nd$^{3+}$ ions ($^4I_{9/2}$) are well separated along the crystallographic $c$ axis, making the interlayer -O-Ta-O-Ta-O- exchange bridges (represented by double-coloured bonds in {\bf a} for two selected sites) much longer and therefore less effective than the intralayer bridges -O-Ta-O- (thin double-coloured bonds in {\bf b}).
{\bf c}, Within the CEF ground-state Kramers' doublet the corresponding $S=1/2$ effective spins (arrows) are dominantly antiferromagnetically coupled to nearest neighbours  within layers (thick single-color bonds), with the exchange interaction being characterized by strong Ising anisotropy along the crystallographic $c$ axis, ${\cal J}_z=0.90(2)\,{\rm K} \gg {\cal J}_{xy}=0.16(2)\,{\rm K}$.
This leads to a quantum spin-liquid ground state with no long-range magnetic order and characteristic low-frequency fluctuations persisting at least down to 66\,mK. 
}
\label{fig1}
\end{figure*}
%



The hexagonal crystal structure (space group $P\bar{6}c2$) of the neodymium heptatantalate NdTa$_7$O$_{19}$ has been known for some time, yet its physical properties have been so far investigated only regarding applications in laser technology and non-linear optics \cite{schaffrath1990chemischen, leonyuk2007new}.
Here we highlight the perfect triangular symmetry of well-separated layers of magnetic Nd$^{3+}$ ions in this material (Fig.\,\ref{fig1}a), representing a highly sought-after setting for realization of the two-dimensional TAFM.
We investigate its magnetism  by performing various complementary experimental techniques on high-quality polycrystalline samples. 
The details on sample synthesis and crystal structure are given in Methods.

\begin{figure*}[t]
\centering
\includegraphics[width=0.9\linewidth]{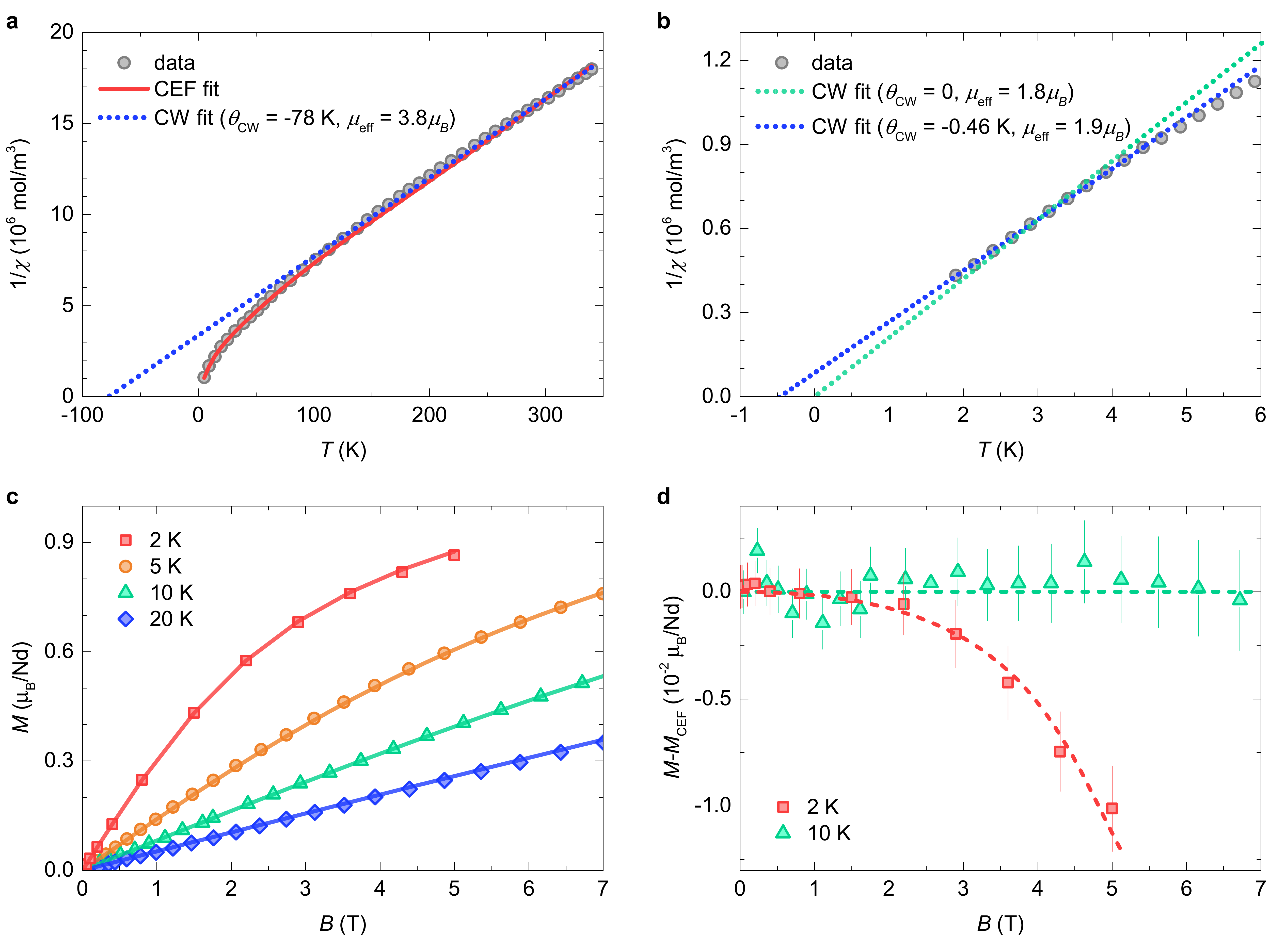}
\caption{{\bf Bulk magnetic properties of NdTa$_7$O$_{19}$:}  
{\bf a,b}, Inverse magnetic susceptibility (without diamagnetic contribution; see Methods) in a magnetic field of 100\,mT suggests absence of any magnetic instabilities in \Nd~al least down to 2~K. 
The dashed lines correspond to the Curie-Weiss (CW) model and indicate small exchange interactions of the order of a kelvin in the crystal-electric-field (CEF) ground state (see {\bf b}).
This agrees well with an excellent fit of a non-interacting CEF model (see Methods) to the magnetic susceptibility in the whole temperature range (solid curve in {\bf a}).
{\bf c,d}, The CEF model (solid lines) also explains well the magnetic-field dependence of magnetization measured at several different temperatures.
Deviations from this model are found only at the lowest temperature of 2\,K and are highlighted in {\bf d}, where the dotted lines serve as a guide to the eye to emphasize the difference between the experimental magnetization and the CEF model.
Error bars represent an uncertainty of one standard deviation.
}
\label{fig2}
\end{figure*}

Bulk magnetic measurements (see Methods) show no sign of magnetic ordering or freezing down to 2\,K (Fig.\,\ref{fig2}), suggesting a possible dynamical magnetic ground state of \Nd.
At temperatures above 100\,K, the inverse magnetic susceptibility $1/\chi$ follows the Curie-Weiss (CW) law (Fig.\,\ref{fig2}a; see Methods) with the Weiss temperature $\theta_{\rm CW}^h = -78(4)$\,K and an effective magnetic moment $\mu_{\rm eff}^h=3.8(1)\mu_B$ of the Nd$^{3+}$ ion, which is close to the free-ion value of $3.62\mu_B$, where $\mu_B$ is the Bohr magneton.
The rather large value of $\theta_{\rm CW}^h$ is not a sign of strong exchange interactions present in \Nd. This is because the excited CEF levels of RE ions typically lie close to the ground state and can strongly affect their magnetism, leading to situations with seemingly strong magnetic interactions even when these are in fact much weaker \cite{simonet2008hidden}.
Indeed, after deviating from the high-temperature CW dependence below 100\,K, a CW dependence is re-established below 6\,K, but with drastically reduced $\theta_{\rm CW}^l = -0.61(5)$\,K and $\mu_{\rm eff}^l=1.9(1)\mu_B$ (Fig.\,\ref{fig2}b).    
\begin{figure*}[t]
\centering
\includegraphics[width=0.9\linewidth]{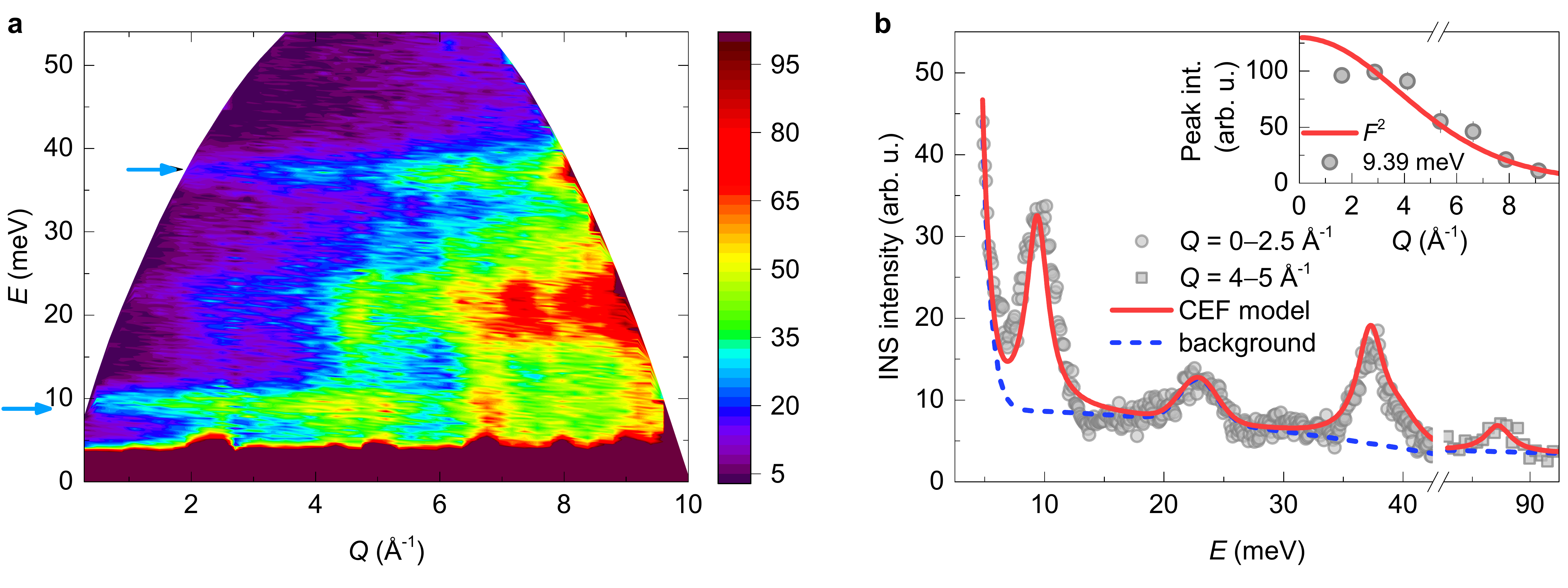}
\caption{{\bf Magnetic and phonon excitations in \Nd}: {\bf a}, INS intensity at 5\,K for the neutron incident energy of 60\,meV reveals the predominance of phonon excitations at larger wave-vector values $Q$ as well as additional CEF magnetic excitations at low $Q$ values, as indicated by arrows.
{\bf b}, Constant-$Q$ cuts (symbols) obtained by integrating over the ranges 0--2.5\,\AA$^{-1}$ and 4--5\,\AA$^{-1}$ at low and high energy transfers, respectively, are compared to the CEF model including background (solid line). 
The background (dashed line) is fitted by three Gaussian functions; two centred at $E$\,=\,0 (a narrow one accounts for elastic scattering while a broad one originates from phonons) and another one at $\sim$23\,meV that is attributed to phonons. 
The inset shows the intensity of the 9.4\,meV CEF peak as a function of $Q$  integrated over 1.25\,\AA$^{-1}$-broad intervals and the magnetic form factor (solid line).}
\label{fig3}
\end{figure*}

To understand the magnetism of \Nd~and quantitatively assess its magnetic interactions, the knowledge of the Nd$^{3+}$ CEF levels is first required. 
For this purpose the dynamical structure factor $S(E,\bf{Q})$ shown in Fig.\,\ref{fig3}a was measured by inelastic neutron scattering (INS; see Methods), which directly detects the CEF transitions.
The INS intensity close to $E=0$ is due to elastic scattering, while the scattering at $E\gtrsim5$\,meV is dominantly due to phonon excitations, as it increases with increasing $Q$ as well as exhibits typical phonon dispersion arcs extending up to $\sim$40\,meV. 
In addition, we find three very flat excitation bands, two stronger at 9.4 and 38\,meV (highlighted by arrows in Fig.\,\ref{fig3}a), and one weaker at 87\,meV (see Supplementary Fig.\,2), which we attribute to CEF transitions.
In contrast to the phonon modes, the CEF modes become weaker with increasing $Q$ due to decreasing magnetic form factor $F^2(\bf{Q})$ of Nd$^{3+}$ ions (inset in Fig.\,\ref{fig3}b).


The composition of the CEF states and their corresponding magnetic characteristics follow from a simultaneous CEF fit (see Methods) of the INS data (Fig.\,\ref{fig3}b), the temperature dependence of the magnetic susceptibility (Fig.\,\ref{fig2}a), and magnetization data at various temperatures (Fig.\,\ref{fig2}c).
The CEF fit predicts that two energy levels almost coincide (see Table\,\ref{CEF-vec}), effectively yielding three CEF transitions between the ground state Kramers' doublet and the four excited Kramers' doublets of the $^4I_{9/2}$ Nd$^{3+}$ multiplet split by the CEF, as observed in the INS experiment.
The first CEF excited doublet lies $\Delta_{\text{CEF}}=9.4$\,meV above the ground state, therefore at temperatures $T\ll \Delta_{\text{CEF}}/k_B = 109$\,K, \Nd~can be considered as an effective spin-1/2 system.
The magnetic character of the derived CEF ground state is highly anisotropic, which reflects in anisotropic $g$ factors $g_{z} =2.78$ and $g_{xy} =1.22$ (equation\,(\ref{g-factor})) for the parallel and perpendicular direction with respect to the crystallographic $c$ axis, respectively.
This sets a magnetic easy-axis perpendicular to the triangular planes,  with the projection of the magnetic moment on the easy axis $\mu_{z}=g_{z}\mu_B/2=1.39\mu_B$ being rather small.
This explains the large reduction of the effective magnetic moment compared to the full free-ion value, which we obtain from the low-temperature CW fit of the magnetic susceptibility (Fig.\,\ref{fig2}b).
As the agreement of the non-interacting CEF model with bulk magnetization data is very good even for $T\ll\Delta_{\text{CEF}}/k_B$ (Fig.\,\ref{fig2}c), the magnetism of NdTa$_7$O$_{19}$ is predominantly determined by CEF effects.
So, any exchange interactions between effective spins $S=1/2$ in the CEF ground state must be small.
This is in line with $\theta_{\rm CW}^l = -0.61(5)$\,K, which suggests the presence of antiferromagnetic exchange interactions of the order of a kelvin.  
We note that the pure non-interacting Curie model with $\theta_{\rm CW} =0$ fits the low-temperature experimental susceptibility significantly worse (Fig.\,\ref{fig2}b), which corroborates with the fact that the non-interacting CEF model cannot perfectly describe the experimental magnetization curves at low temperatures, e.g., at 2\,K in Fig.\,\ref{fig2}d.  
\begin{figure*}[!]
\centering
\includegraphics[width=0.9\linewidth]{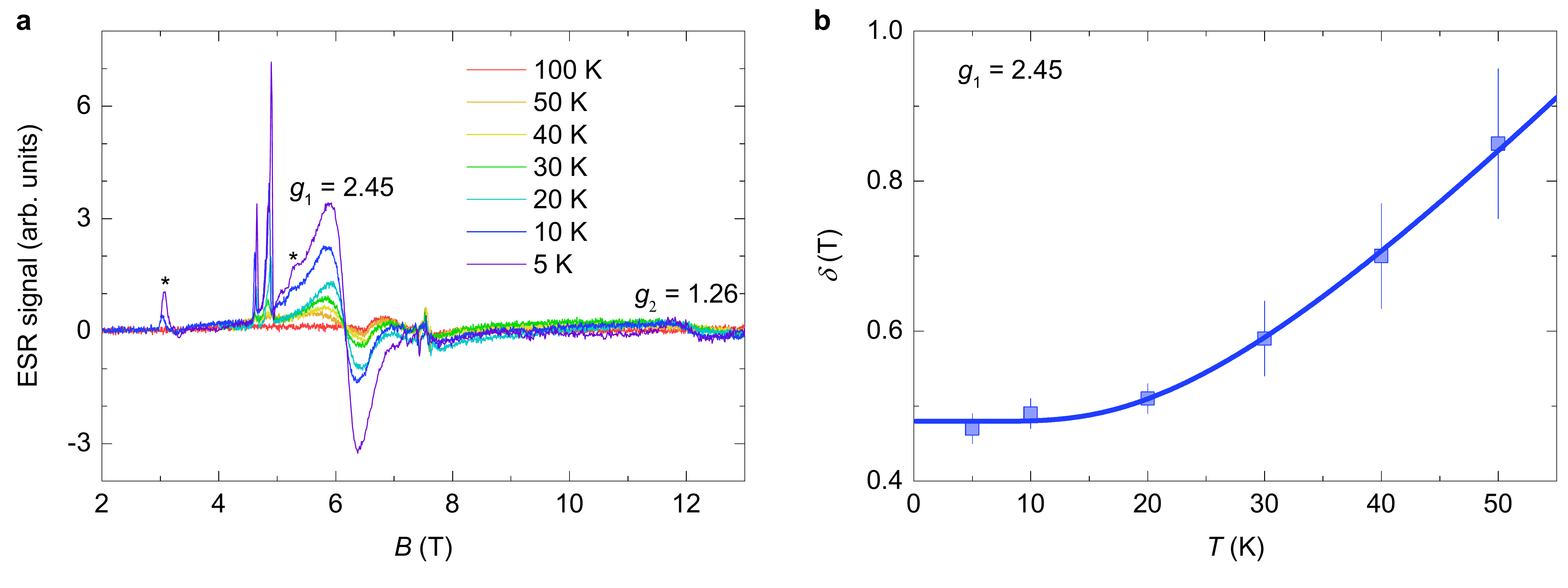}
\caption{{\bf Determination of magnetic anisotropy in \Nd}: {\bf a},  ESR spectra recorded at 211.9\,GHz show two dominant lines at $g$ factors $g_1=2.45$ and $g_2=1.26$ that correspond well to the values $g_z=2.78$ and $g_{xy}=1.22$ obtained from CEF modelling.
The asterisks mark signals of oxygen impurities. 
A few additional narrow impurity signals are detected, which however have orders of magnitude smaller intensities than the main broad lines.
{\bf b}, The temperature dependence of the peak-to-peak ESR linewidth of the main signal at $g_1$ (symbols) is explained by the Orbach model (line) of equation\,(\ref{OrbsachESR}), while the large low-temperature saturation value of the linewidth is accounted for by a large exchange anisotropy.
Error bars represent an uncertainty of one standard deviation.}
\label{fig4}
\end{figure*}
%

 
\begin{figure*}[t]
\centering
\includegraphics[width=0.9\linewidth]{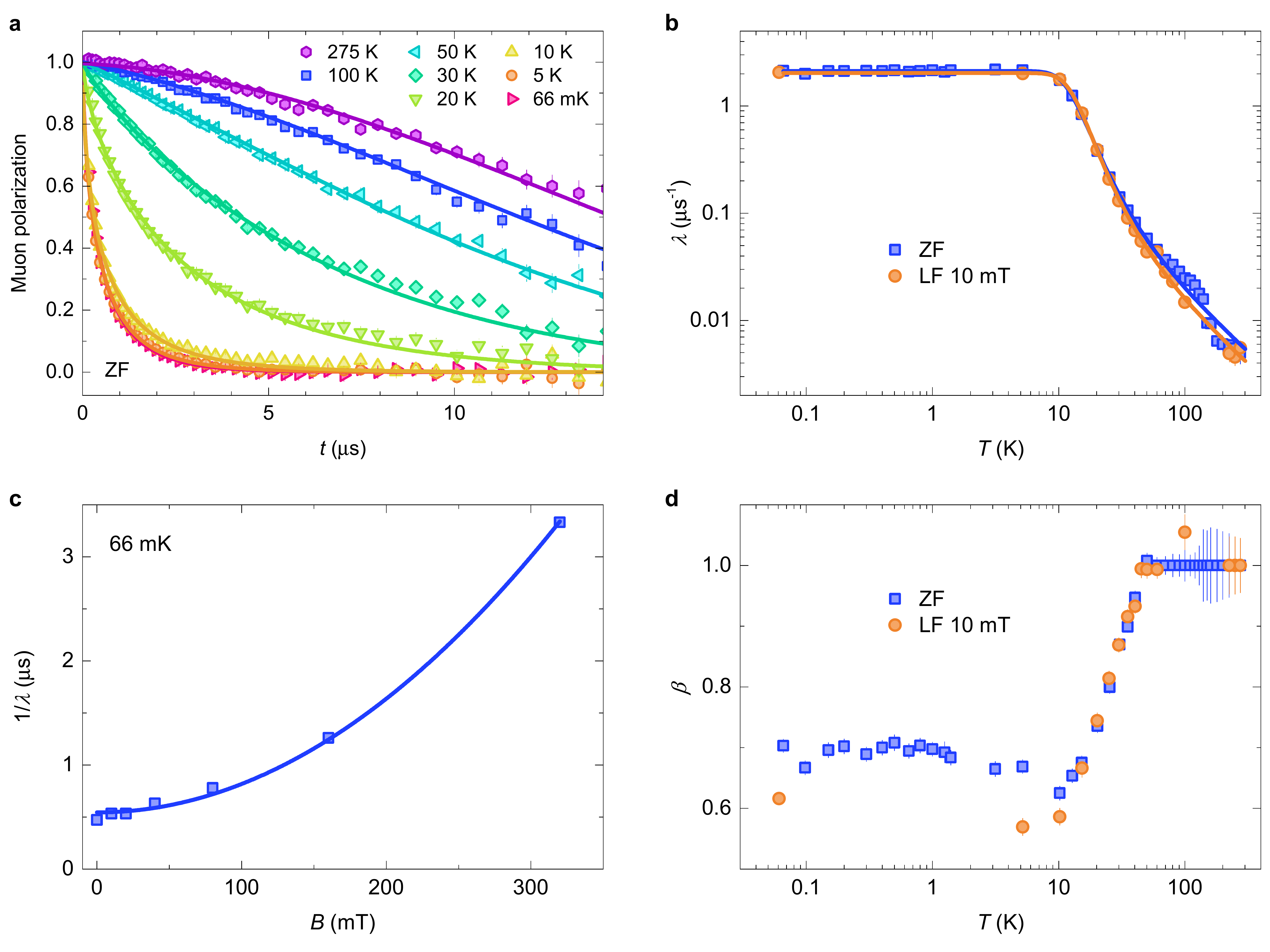}
\caption{{\bf Quantum SL ground state of \Nd~evidenced by $\mu$SR}: {\bf a}, Monotonic muon-polarization decay is observed in zero applied magnetic field (ZF) at all temperatures, which reveals absence of any magnetic ordering or freezing down to at least 66\,mK. 
The solid lines correspond to fits with equation\,(\ref{musrFunc}). 
{\bf b,d}, The muon spin relaxation rates $\lambda$ due to dynamical local fields (see {\bf b}) and corresponding stretching exponents $\beta$ (see {\bf d}) are obtained from these fits for ZF, as well as for a small longitudinal field (LF) of 10\,mT. 
The temperature dependence of $\lambda$ follows the Orbach model of equation\,(\ref{OrbachmuSR}). 
{\bf c}, The muon spin relaxation rate measured at 66\,mK as a function of the longitudinal magnetic field follows the Redfield model of equation\,(\ref{Red}) corresponding to fast spin fluctuations.
Error bars represent an uncertainty of one standard deviation.}
\label{fig5}
\end{figure*}

A more quantitative insight into exchange interactions in the ground-state Kramers' doublet of \Nd~is obtained from electron spin resonance (ESR) measurements (see Methods).
This method directly measures $g$ factors and thus also serves to verify the predictions of the CEF model.
The center of the main broad ESR signal corresponds to $g$ factor of $g_1$\,=\,2.45, while a substantially weaker signal with similar linewidth is found at $g_2$\,=\,1.26 (Fig.\,\ref{fig4}a).
Considering that the $g$ factors of Nd$^{3+}$ ions depend dramatically on the CEF parameters \cite{abragam1970electron},
the positions of these two signals are in rather good agreement with the values predicted by our CEF modelling, $g_{z}=2.78$ and $g_{xy}=1.22$.
The ESR linewidth $\delta$ of the main signal exhibits a pronounced temperature dependence and saturates at a constant value $\delta_0=0.48(1)$\,T below $\sim$10\,K (Fig.\,\ref{fig4}b).
This temperature variation is attributed to the Orbach mechanism (see Methods) accounting for two-phonon scattering via excited CEF levels and shows a gapped behaviour with a gap $\Delta_{\rm ESR}$ expected to be of the order of the CEF gap to the lowest excited Kramers' doublet \cite{abragam1970electron}.
Indeed, we find a reasonably good agreement between $\Delta_{\rm ESR}=6.5(5)$\,meV and $\Delta_{\text{CEF}}=9.4$\,meV obtained from INS measurements.

The low-temperature saturation value of the ESR linewidth $\delta_0=0.48(1)$\,T is used to assess the exchange interactions and their anisotropy in the CEF ground state.
Such a large linewidth can only be explained by strong exchange anisotropy, which is indeed expected for RE-based magnets \cite{abragam1970electron}.
The broadening due to dipolar interactions, on the other hand, amounts to only about $\sim$5\,mT and is thus negligible. 
The exchange anisotropy arises from projecting the intrinsic isotropic exchange interaction ${\cal J}_0$ between spin degrees of freedom onto the effective $S=1/2$ spin-orbit entangled degrees of freedom of the ground-state Kramers' doublet, leading to exchange-interaction components \cite{abragam1970electron}
${\cal J}_{\alpha}={\cal J}_0g_{\alpha}^2 (g_J-1)^2/g_J^2$,
where $\alpha=z,\,xy$ and $g_J=8/11$ is the Land\'e factor of Nd$^{3+}$.
The strong uniaxial anisotropy of the $g$ factor is thus reflected in a strongly anisotropic Ising-like spin Hamiltonian 
\begin{equation}
\begin{split}
\mathcal{H}_{\rm ex} 
& = \sum_{\langle i,j\rangle} {\cal J}_z S_i^z S_j^z + {\cal J}_{xy} \left( S_i^x S_j^x + S_i^yS_j^y\right) \\
& = \sum_{\langle i,j\rangle}  \left[ \left({\cal J} - \tfrac{1}{2}D \right) \left( {\bf S}_i \cdot {\bf S}_j \right) +  \tfrac{3}{2}D  S_i^zS_j^z \right],
\label{eqHam}
\end{split}
\end{equation}
which is used to describe the magnetism of the CEF ground state.
Here the sum runs over the nearest-neighbour spin pairs on the triangular lattice, ${\cal J}$ is the isotropic part of the exchange interaction and $D$ is the anisotropic part described by a traceless tensor, yielding ${\cal J}_z$\,=\,${\cal J}+D$ and ${\cal J}_{xy}$\,=\,${\cal J}-D/2$. 
The anisotropic exchange interaction is evaluated as $D$\,=\,$\sqrt{8/27}g_{z}\mu_B\delta_0/k_B=0.49(1)$\,K (see equation\,(\ref{ESRlinewidth})), which then yields the isotropic exchange interaction ${\cal J}=D\big(g_{z}^2/2+g_{xy}^2\big)\big/\big(g_{z}^2-g_{xy}^2\big)=0.86D=0.41(1)$\,K.
\Nd~is thus a strongly anisotropic spin-1/2 Ising-like TAFM, with ${\cal J}_z=0.90(2)$\,K and ${\cal J}_{xy}=0.16(2)$\,K.
The magnitude of the estimated exchange interactions predicts the powder-averaged Curie-Weiss temperature $\theta_{\rm CW} = -3{\bar{\cal J}}/2=-0.62(2)$\,K, where ${\bar{\cal J}}=({\cal J}_z+2{\cal J}_{xy})/3$ is the average exchange constant.  
This agrees perfectly with the low-temperature CW temperature $\theta_{\rm CW}^l = -0.61(5)$\,K determined from bulk magnetic susceptibility.


Since the magnetic interactions in \Nd~are rather small, investigation of its magnetic ground state requires measurements in millikelvin temperature range.
Here we resort to muon spin relaxation ($\mu$SR) spectroscopy (see Methods), a technique that provides a very sensitive local probe that can detect magnetic fields as small as those originating from nuclear magnetic moments \cite{yaouanc2011muon}.
These measurements reveal a magnetically disordered ground state.
Namely, down to the lowest experimental temperature of $66\,{\text mK}\ll {\cal J}_z$ the muon polarization remains a monotonically decaying function (Fig.\,\ref{fig5}a), which is a hallmark of dynamical local fields of electronic origin present at the muon stopping site \cite{yaouanc2011muon}.
Static internal magnetic fields, in contrast, would result in coherent oscillations of the muon polarization.
The muon spin relaxation rate $\lambda$ due to dynamical local fields (see Methods) remains constant up to around 10\,K and then starts decreasing profoundly with increasing temperature (Fig.\,\ref{fig5}b).
Such temperature dependence corresponds to the presence of two relaxation mechanisms -- a temperature-independent one, characteristic of the CEF ground-state Kramers' doublet, and a thermally-activated one that prevails at higher temperatures.
As with the ESR linewidth, we find that the temperature dependence of $\lambda$ is due to the Orbach process given by equation\,(\ref{OrbachmuSR}), yielding the energy gap $\Delta_{\mu\text{SR}}=5.7(5)$\,meV, which is in very good agreement with the estimate $\Delta_{\rm ESR}=6.5(5)$\,meV from ESR. 
We note that a very similar behaviour with a low-temperature relaxation plateau was found also in Nd-based langasite featuring a frustrated kagome lattice \cite{zorko2008easy}, as well as in a few other RE-based frustrated magnets \cite{gardner1999cooperative, li2016muon, clark2019two}, and is typically regarded as a fingerprint of a correlated disordered ground state. 

A further insight into spin fluctuations characteristic of the disordered magnetic ground state of \Nd~is provided by the field dependence of the muon spin relaxation rate at the plateau. 
The measurements performed at 66\,mK (Fig.\,\ref{fig5}c) show that the system is in the fast-fluctuation regime as $\lambda$ obeys the Redfield relation of equation\,(\ref{Red}). 
The fit to the data yields the spread of local fields at the muon stopping site $B_{\text{loc}}=12(1)$\,mT and the corresponding spin fluctuation frequency $\nu_e=120(10)$\,MHz.
While the derived $B_{\text{loc}}$ is comparable to typical values of fluctuating magnetic fields at muon stopping sites in insulating oxides \cite{yaouanc2011muon}, $\nu_e$ is very small compared to the spin fluctuation frequency expected in a paramagnetic state $k_B \bar{\cal J}/\hbar \simeq 50$\,GHz.
This is most likely due to emergent spin excitations related to larger correlated spin-loop structures of frustrated lattices, for which slow dynamics is expected \cite{yaouanc2015evidence}, and thus implies a correlated magnetic ground state of \Nd.
As this state is characterized by strong magnetic anisotropy, strongly orientation-dependent decay of temporal spin correlations should reflect in a broad distribution of muon spin relaxation rates in polycrystalline sample.
Indeed, this is observed experimentally via a pronounced decrease of the stretching exponent $\beta$ (Fig.\,\ref{fig5}d) characterizing the decay of the muon polarization (see Methods), from the high-temperature value of 1 to 0.68(3) at low temperatures. 
This corresponds to strong widening of the distribution of $\lambda$ in the magnetic ground state. 

Having established \Nd~as a new RE-based TAFM compound with a spin-liquid ground state, we next compare it to the most intensively investigated example of this model -- the YbMgGaO$_4$ compound.
We first note that the two systems behave very similarly. 
Just as in \Nd, also in YbMgGaO$_4$ the muon spin relaxation rate exhibits a low-temperature plateau and is characterized by a decrease in the stretching exponent when entering the low-temperature region \cite{li2016muon}. 
Furthermore, the CW temperature of the latter compound is also of the order of 1\,K and the ESR linewidth is very similar to the one we found in \Nd~\cite{li2015rare}, suggesting that the two systems have similarly strong exchange anisotropies.
Nevertheless, we stress that in NdTa$_7$O$_{19}$ the dominant anisotropy is of the Ising type with ${\cal J}_z/{\cal J}_{xy}$\,=\,5.6, while in YbMgGaO$_4$ it is of the easy-plane type with ${\cal J}_z/{\cal J}_{xy}$\,=\,0.5 \cite{li2015rare}.
The fact that \Nd~is found in the region of the TAFM phase diagram that is very close to the pure Ising limit is remarkable, as it contrasts any other well-studied compound.

The magnetically disordered ground state of \Nd~is thus attributed to the strong Ising character of the dominant exchange interaction and is observed in absence of any structural disorder, contrary to previous cases \cite{zhu2017disorder, kimchi2018valence, watanabe2014quantum, kawamura2019nature}.
The Ising interaction ${\cal J}_z$ acts between the effective spin-1/2 spin-orbit entangled degrees of freedom within the CEF ground-state Kramers' doublet.
The observed low-frequency excitations, persisting to the lowest temperatures, reveal a potentially crucial role of an additional non-commuting exchange interaction ${\cal J}_{xy}$ that lifts the degeneracy of the classical Ising model and introduces quantum dynamics \cite{moessner2001ising}, which can be understood within the famous resonating-valence-bond scenario suggested by Fazekas and Anderson \cite{anderson1973resonating,fazekas1974ground}.
A similar mechanism is responsible for stabilization of a quantum SL phase of pyrochlores known as quantum spin ice \cite{savary2017quantum,gingras2014quantum}
and thus appears to be universal among various highly anisotropic frustrated spin lattices. 
Our results open a new perspective for modern theoretical approaches, which have so far focused only on unconventional phases in the easy-plane region of the TAFM phase diagram \cite{zhu2018topography, maksimov2019anisotropic}.
The effect of anisotropy in the TAFM model could be investigated experimentally more thoroughly by inspecting various members of the large family of rare-earth heptatantalates and other related RE-based TAFM compounds. 
These quantum materials provide an optimal framework for realizations of quantum spin liquids borne out of magnetic anisotropy, similarly as observed in the paradigmatic Kitaev honeycomb model \cite{kitaev2006anyons}. 
Such states are also of high technological relevance, as they represent a promising platform for quantum computing \cite{nayak2008non}.  

\section{References}

\section{Methods}
{\bf Synthesis and crystal structure.}
Polycrystalline sample of \Nd~was prepared by conventional solid state synthesis method from high purity Nd$_2$O$_3$ (Alfa Aesar, 99.994\%) and Ta$_2$O$_5$ (Alfa Aesar, 99.993\%) \cite{zuev1991x}.
Prior to use, Nd$_2$O$_3$ was preheated at 900$^\circ$C for 12\,h. 
Stoichiometric amounts of each material were mixed and pelletized. 
Single phase \Nd~sample was obtained after heat treatment of the reactants for several days at 950$^\circ$C, 1000$^\circ$C, 1050$^\circ$C, 1100$^\circ$C, 1150$^\circ$C, and 1200$^\circ$C with intermediate grindings. 
Its phase purity was checked with Rigaku X-Ray diffractometer at room
temperature using Cu $K_\alpha$ radiation.
The Rietveld refinement was carried out within the space group $P\bar{6}c2$ using {\sc fullprof suite} \cite{rodriguez1993recent}.
The results are shown in Supplementary Fig.\,1 and the parameters are summarized in Supplementary Table\,I. 
No anti-site disorder was detected in this compound.

The magnetic Nd$^{3+}$ ions occupy the $2c$ Wyckoff position in the unit cell and form an equilateral triangular lattice within the $ab$ crystal plane (Fig.\,\ref{fig1}). 
The nearest neighbour distance between the magnetic ions is 6.224\,\AA~
and is much smaller than the inter-layer distance of $c/2=9.969$\,\AA.
Furthermore, the exchange interaction within the triangular layers is mediated via -O-Ta-O- exchange bridges (Fig.\,\ref{fig1}b), while -O-Ta-O-Ta-O- bridges that are involved in the interlayer exchange (Fig.\,\ref{fig1}a) are much longer.
Therefore, the deviation of the spin lattice in \Nd~from the two-dimensional TAFM model is expected to be very small.  

{\bf Bulk magnetic measurements.} Bulk magnetic measurements were performed on a polycrystalline \Nd~sample enclosed in a plastic container.
Quantum Design SQUID magnetometers were employed in the temperature range between 2 and 340\,K and field range up to 7\,T.
The magnetic susceptibility $\chi=M/H$, where $M$ is sample magnetization, was measured in a magnetic field $\mu_0 H = 100$\,mT,  with $\mu_0$ being vacuum permeability.
The susceptibility of the sample container was measured separately and subtracted from the total susceptibility.
Also the diamagnetic susceptibility of the \Nd~sample $\chi_{\rm dia}=-4.3\times 10^{-3}$\,cm$^3$/mol \cite{bain2008diamagnetic} was subtracted to obtain the paramagnetic susceptibility  shown in Fig.\,\ref{fig2}.   

Thus obtained $\chi$ was compared to the Curie-Weiss (CW) model $\chi=C/(T-\theta_{\rm CW})$ in different temperature regions, where $\theta_{\rm CW}$ is the CW temperature and the Curie constant is given by $C=N_A\mu_0\mu_{\rm eff}^2/(3k_B)$, with $N_A$, $k_B$, and $\mu_{\rm eff}$ being the Avogadro number, the Boltzmann constant, and an effective magnetic moment, respectively.
The CW fit in the high-temperature range above 100\,K (Fig.\,\ref{fig2}a) yields the CW temperature $\theta_{\rm CW}^h=-78(4)$\,K and the effective moment $\mu_{\rm eff}^h = 3.8(1)\mu_B$.
The latter is close to the value $g_J\sqrt{J(J+1)}\mu_B=3.62\mu_B$ expected for a free $^4I_{9/2}$ Nd$^{3+}$ ion.
The fit with the same model in the low-temperature range below 6\,K (Fig.\,\ref{fig2}b) gives $\theta_{\rm CW}^l=-0.61(5)$\,K and $\mu_{\rm eff}^l = 1.9(1)\mu_B$.

{\bf Inelastic neutron scattering.} INS measurements were performed using the MARI spectrometer at the ISIS Pulsed Neutron and Muon Source of the Rutherford Appleton Laboratory, UK. 
The direct geometry of the spectrometer has a continuous detector coverage from 3.5–135$^\circ$.
Approximately 5.3\,g of polycrystalline sample was loaded into an aluminium sample holder with an annular geometry.
The dynamical structure factor $S(E,\bf{Q})$, with $\bf{Q}$ being the scattering wave vector, was measured at 5\,K for the incident energies $E=60$ and 100\,meV, selected using a Fermi chopper system with a Gd foil chopper pack rotating at 200 Hz.
The chosen configuration of the instrument ensured an elastic line resolution of $\delta E/E$\,$\approx$\,0.04 and $Q$ ranges $\sim$0.25–9.5\,\AA$^{-1}$ and $\sim$0.35–12.7\,\AA$^{-1}$ for $E=60$ and 100\,meV, respectively.
We note that the magnetic signal of the CEF modes is rather weak compared to the signal of the phonon modes (Fig.\,\ref{fig3}a and  Supplementary Fig.\,2), because the fraction of the magnetic Nd$^{3+}$ ions with respect to all ions in \Nd~is only 1:27.
Moreover, the highest CEF mode at 87(1)\,meV is barely noticeable, because the data at this energy start only at $Q=3.5$\,\AA$^{-1}$ (see Supplementary Fig.\,2), where the magnetic form factor is already significantly reduced.

{\bf CEF modelling.}
\begin{table}[b]
\caption{CEF parameters (in meV) obtained from a point-charge (PC) model and derived from the fit of the INS and bulk magnetic data with the CEF Hamiltonian $\mathcal{H}_{\text{CEF}}$.
\label{CEF-par}}
\begin{ruledtabular}
\begin{tabular}{c|c|c|c|c|c|c}
 & $B_2^0$ & $B_4^0$ & $B_4^3$  & $B_6^0$  & $B_6^3$  & $B_6^6$  \\
\hline
PC & -0.0536  &  0.00090 &  0.11091  &  -0.00016 & -0.00081  & 0.00024\\
\hline
fit & 0.0417 & -0.00003 & -0.1515 & -0.00046 & 0.00390 & -0.00438 \\
\end{tabular}
\end{ruledtabular}
\end{table}
\begin{table}[t]
\caption{The eigenstates $\pm\omega_k$ ($k$\,=\,0-4) of the CEF Hamiltonian given in the $|\pm m_J\rangle$ basis and the corresponding energies of the $^4I_{9/2}$ Nd$^{3+}$ multiplet in \Nd.
\label{CEF-vec}}
\begin{ruledtabular}
\begin{tabular}{c|c|c|c|c|c}
$|\pm m_J\rangle$  & $\pm\omega_{0}$ & $\pm\omega_{1}$ & $\pm\omega_{2}$ & $\pm\omega_{3}$ & $\pm\omega_{4}$ \\
               \hline
$|\pm9/2\rangle$ & 0           & $\pm$0.590  & 0          & $\pm$0.807 & 0          \\
$|\pm7/2\rangle$ & 0           & 0           & $\pm$0.425 & 0          & 0          \\
$|\pm5/2\rangle$ & 0.933       & 0           & $-$0.017   & 0          & $\mp$0.358  \\
$|\pm3/2\rangle$ & 0           & 0.021       & 0          & $-$0.015   & 0          \\
$|\pm1/2\rangle$ & 0           & 0           & $-$0.515   & 0          & 0          \\
$|\mp1/2\rangle$ & $\mp$0.244  & 0           & $\mp$0.574 & 0          & $-$0.588   \\
$|\mp3/2\rangle$ & 0           & $\pm$0.807  & 0          & $\mp$0.590 & 0          \\
$|\mp5/2\rangle$ & 0           & 0           & $\pm$0.015  & 0          & 0          \\
$|\mp7/2\rangle$ & 0.263       & 0           & $-$0.474   & 0          & $\pm$0.725 \\
$|\mp9/2\rangle$ & 0           & 0           & 0          & 0          & 0          \\
\hline
$E$(meV)  &    0    &   9.4      & 37.3        & 39.8        & 87.1            
\end{tabular}
\end{ruledtabular}
\end{table}
The five CEF Kramers' doublets can be written as linear combinations of the eigenstates $|\pm m_J\rangle$ of the $z$-component of the total-spin operator, where $m_J$\,=\,$(2n-1)/2$, $n$\,=\,1-5.
The composition of these states depends on the CEF Hamiltonian
$\mathcal{H}_{\text{CEF}} = \sum_{l,m} B_l^m O_l^m$,
where $O_l^m$ are Stevens operators \cite{stevens1952matrix} and $B_l^m$ are the corresponding scaling parameters. 
In order to determine these parameters, allowed by the Nd$^{3+}$-site point symmetry in NdTa$_7$O$_{19}$ with a dihedral $D_3$ group, we simultaneously fitted the INS data (Fig.\,\ref{fig3}b), magnetic-susceptibility data (Fig.\,\ref{fig2}a) and magnetization data (Fig.\,\ref{fig2}c) using Mantid software \cite{arnold2014mantid}.
The combined fit agrees well with all experiments and yields the $B_l^m$ parameters summarized in Table\,\ref{CEF-par}.
We note that the parameters significantly deviate from those predicted by a point-charge model (Table\,\ref{CEF-par}), as often found in RE-based systems \cite{newman2007crystal}.

The estimated $B_l^m$ parameters predict the five Kramers' doublets given in Table\,\ref{CEF-vec}.
All four transitions from the CEF ground state to the excited states occur for energy transfers below 100\,meV, which is covered by our experiments.
However, only three flat INS modes are observed experimentally, because the energies of two excited states almost coincide and yield a single broad peak at 38 meV.
Knowing the composition of the CEF ground state allows us to determine the corresponding $g$ factor anisotropy \cite{abragam1970electron},
\begin{equation}
\begin{split}
g_{z} = 2g_J |\langle \pm\omega_0|J_z|\pm\omega_0\rangle| = 2.78, \\
g_{xy} = g_J |\langle \pm\omega_0|J_\pm|\mp\omega_0\rangle| =1.22,
\label{g-factor}
\end{split}
\end{equation}
where ${z}$ and ${xy}$ denote directions parallel and perpendicular to the high-symmetry $c$ axis that is perpendicular to the triangular layers.

{\bf Electron spin resonance.} ESR measurements were performed at the National High Magnetic Field Laboratory, Tallahassee, USA on a custom-made transmission-type ESR spectrometer with homodyne detection equipped with a sweepable 15-T superconducting magnet.
The measurements were performed in the Faraday configuration at the irradiation frequency of 211.9\,GHz on a 200-mg polycrystalline sample in a Teflon container.
The temperature was controlled using a continuous-flow He cryostat in the temperature range between 5 and 100\,K.
A standard field-modulation technique was used with the modulation field of about 2\,mT to enhance the signal-to-noise ratio.

The temperature dependence of the ESR linewidth (Fig.\,\ref{fig4}b) obeys the relation 
\begin{equation}
\delta(T)=\delta_0 + \frac{a}{{\rm exp}(\Delta_{\rm ESR}/T)-1}.
\label{OrbsachESR}
\end{equation}
The constant $\delta_0$ term is attributed to magnetic anisotropy in the ground-state Kramers' doublet while the exponential term describes the Orbach relaxation process, arising from CEF fluctuations due to two-phonon scattering via intermediate CEF states \cite{abragam1970electron}.
In this process, the spin-fluctuations frequency dramatically increases with temperature due to increasing phonon density,
\begin{equation}
\nu_e \propto 1/ \left[ \exp\left(\Delta/k_BT\right)-1\right]
\label{elfreq}
\end{equation}
causing ESR line broadening. 
The fit to the experimental data yields the energy gap $\Delta_{\rm ESR}= 6.5(5)$\,meV, the scaling constant $a=1.3(1)$\,T and the zero-temperature linewidth $\delta_0=0.48(1)$\,T.

In the first approximation, i.e., neglecting exchange-narrowing effects, the zero-temperature ESR linewidth is given by
\begin{equation}
\delta_0=\sqrt{M_2}/g\mu_B,
\label{ESRlinewidth}
\end{equation}
where $M_2=\frac{1}{3}zS(S+1)(3D/2)^2$ is the second moment of the absorption line \cite{abragam1970electron} and $z$\,=\,6 denotes the number of nearest neighbours on the triangular lattice.
We stress that the estimated exchange anisotropy is stronger than the isotropic exchange interaction, $D>J$ (see main text), which justifies omission of the exchange-narrowing effects \cite{abragam1970electron}.

{\bf Muon spin relaxation.} $\mu$SR measurements were performed on the MUSR spectrometer at the ISIS Pulsed Neutron and Muon Source of the Rutherford Appleton Laboratory, UK.
About 1\,g of polycrystalline sample was put on a silver sample holder and was covered by diluted GE varnish to ensure good thermal conductivity at millikelvin temperatures.
The experiment was performed in zero field and in various longitudinal fields in the temperature range between 66\,mK and 275\,K.
A dilution-refrigerator set-up was used to reach the lowest temperatures, while a standard He-flow cryostat was used for temperatures above 2\,K.
Several runs at the same experimental conditions were done to calibrate the data between the two set-ups.

As muons stopping in the sample are initially almost fully polarized, the time dependence of their polarization $P(t)$ provides an extremely sensitive probe of local magnetic
fields \cite{yaouanc2011muon}.
The experimentally measured muon asymmetry is proportional to the muon polarization,
$A(t) = A_0 P(t) + A_{\rm bgd}$,
 where $A_0$ and $A_{\rm bgd}$ are the initial asymmetry and the background asymmetry, respectively.
The experimental data curves shown in Fig.\,\ref{fig5}a and Supplementary Fig.\,3 are described by the model
\begin{equation}
P(t) = P_{\rm KT} (t,\Delta,B) \exp[-(\lambda t)^\beta],
\label{musrFunc}
\end{equation}
which takes into account both, the muon depolarization due to small disordered static internal fields of nuclear origin with a distribution width $\Delta$, as described by the standard Kubo-Toyabe function $P_{\rm KT} (t,\Delta,B)$ \cite{yaouanc2011muon}, and the polarization decay due to fluctuating local fields of electronic origin, characterized by the stretched exponential function $\exp[-(\lambda t)^\beta]$.
Here $\lambda$ is the corresponding muon spin relaxation rate and the stretching exponent $\beta < 1$ denotes a finite distribution of $\lambda$. 
Weak nuclear magnetic fields and experimental background were determined from the field dependence of $A(t)$ at 275\,K (see Supplementary Fig.\,3), which gives $\Delta=0.062$\,mT, $\lambda = 0.005$\,$\mu$s$^{-1}$, $\beta = 1$, $A_{0}=0.283$ and $A_{\rm bgd}=0.037$ (in the dilution-refrigerator set-up the latter parameters are $A_{0}=0.287$ and $A_{\rm bgd}=0.055$). 
The muon relaxation at high temperatures and low applied fields is predominantly due to static nuclear fields, as the relaxation due to dynamical fields of electronic origin is weaker. Yet the letter remains the only relaxation mechanism once the applied field exceeds 1~mT (Supplementary Fig.\,3).
With lowering temperature, however, the electronic dynamical relaxation mechanism becomes dominant even in zero applied field.
The temperature dependence of $\lambda$ is obtained from fits of equation\,(\ref{musrFunc}) to the experimental $P(t)$ curves (Fig.\,\ref{fig5}a) and is shown in Fig.\,\ref{fig5}b, while the corresponding stretching exponent is shown in Fig.\,\ref{fig5}d.
We note that deviation of $\beta$ from unity at low temperatures can be understood within the scenario of strong exchange anisotropy governing the spin dynamics in the ground-state Kramers' doublet of \Nd, resulting in a broad distribution of local fields at the muon site. 
If, alternatively, $\beta < 1$ originated from muons stopping at multiple sites in the unit cell, this parameter would not change with temperature. 

Assuming that the dynamics of the local fields is well characterized by a single electron-spin fluctuation frequency $\nu_e$, $\lambda$ is inversely proportional to $\nu_e$ in the limit of low applied fields ($\gamma_\mu B_0 \ll \nu_e$), as follows from the Redfield relation \cite{yaouanc2011muon} 
\begin{equation}
\lambda = \frac{2\gamma_\mu^2 B_{\text{loc}}^2 \nu_e}{\gamma_\mu^2 B_0^2+\nu_e^2},
\label{Red}
\end{equation}
valid in the case of fast fluctuations, i.e., for $\nu_e \gg \gamma_\mu B_{\rm loc}$.
Here $\gamma_\mu$\,=\,135.5$\times$2$\pi$\,MHz/T is the muon gyromagnetic ratio and $B_{\rm loc}$ is the width of the distribution of fluctuating local fields.
Indeed, the magnetic-field dependence of $\lambda$ at the lowest temperature of 66\,mK follows this dependence (Fig.\,\ref{fig5}c) and yields the parameters $B_{\text{loc}}=12(1)$\,mT and $\nu_e=120(10)$\,MHz,
which justify the use of the fast-fluctuation model, as $\nu_e \gg \gamma_\mu B_{\rm loc}=1.6$\,MHz.

Combining two independent mechanisms of local-field fluctuations in the fast-fluctuation limit ($\nu_e=\nu_{e,1}+\nu_{e,2}$) then yields  $1/\lambda= 1/\lambda_1+1/\lambda_2$.
Such a model is required to explain the temperature dependence of the muon spin relaxation rate in \Nd~where we find that the form
\begin{equation}
\frac{1}{\lambda}=\frac{1}{\lambda_0} + \frac{\eta}{\exp\left(\Delta_{\mu\text{SR}}\big/T\right)-1},
\label{OrbachmuSR}
\end{equation}
well reproduces the experiment in the whole temperature range (Fig.\,\ref{fig5}b).
Here, the first term is due to spin fluctuations in the CEF ground-state Kramers' doublet, while the second term is due to the Orbach process inducing temperature-dependent spin fluctuations with frequency given by equation\,(\ref{elfreq}) \cite{yaouanc2011muon}, as found also with ESR.
The fit gives $\lambda_0=2.1(1)\,\mu$s$^{-1}$, $\eta=52(8)$\,$\mu$s, and  $\Delta_{\mu\text{SR}}=5.7(5)$\,meV.
We note that single-phonon and Raman-scattering contributions, proportional to $T$ and $T^9$ \cite{abragam1970electron}, respectively, are too steep in the relevant temperature ranges, and are thus not relevant in \Nd.

\section{Data availability}
The data that support the findings of this study are available from the
corresponding author upon reasonable request.

\section{Acknowledgements}
We thank B. D. Gaulin for fruitful discussion. 
The financial support of the Slovenian Research Agency under program No.~P1-0125 and projects No.~J1-2461, No.~BI-US/18-20-064, and No.~N1-0148 is acknowledged.
Also the financial  support from
Science and Engineering Research Board, and Department of Science and Technology, India through research grants ECR/2017/000311
and  DST/INT/RUS/RSF/P-22, respectively, is acknowledged.
The National High Magnetic Field Laboratory is supported by National Science Foundation through NSF/DMR-1644779 and the State of Florida.
INS and $\mu$SR experiments at the ISIS Neutron and Muon Source were supported by beam-time allocations RB1920405 and RB1910518, respectively, from the Science and Technology Facility Council.

This preprint has not undergone peer review or any post-submission improvements or corrections. 
The Version of Record of this article is published in Nature Materials, and is available online at \href{https://doi.org/10.1038/s41563-021-01169-y}{https://doi.org/10.1038/s41563-021-01169-y}.

\section{Author contributions}
T.A. and B.S. contributed equally to this work and are both assigned first authorship of the paper.
P.K. conceived the investigation of \Nd, while A.Z. is the corresponding author who designed this project and supervised the experiments presented in this work.
B.S. and P.K. synthesized and structurally characterized the sample.
Z.J. and B.S. performed bulk magnetic measurements.
T.A., M.P., and M.D.L. conducted the INS measurements, M.P. analysed the results and performed the CEF modelling. 
A.Z. and A.O. performed the ESR measurements and A.Z. analysed the results.
T.A., P.B., and A.Z. conducted the $\mu$SR investigation, T.A. analysed the corresponding data.
All authors discussed the results and the paper.
A.Z. wrote the paper, with inputs provided by M.P. and P.K.
  
\section{Competing interests}
The authors declare no competing interests.
\section{Additional information}
Supplementary information is available for this paper.

Correspondence and requests for materials should be addressed to A.Z.

\newpage
\begin{widetext}
\vspace{19cm}
\begin{center}
{\large {\bf Supplementary Information: Quantum spin liquid in the Ising triangular-lattice antiferromagnet neodymium heptatantalate}}\\
\vspace{0.5cm}
{T. Arh,$^{1,2}$ B. Sana,$^{3}$ M. Pregelj,$^{1}$ P. Khuntia,$^{3}$ Z. Jagli\v{c}i\'c,$^{4,5}$ M. D. Le,$^{6}$ P. Biswas,$^{6}$ A. Ozarowski,$^{7}$ and A.\nolinebreak Zorko$^{1,2,*}$}
\vspace{0.3cm}

{\it \small
$^1$Jo\v{z}ef Stefan Institute, Jamova c.~39, SI-1000 Ljubljana, Slovenia\\
\vspace{0cm}
$^2$Faculty of Mathematics and Physics, University of Ljubljana, Jadranska u. 19, 1000 Ljubljana, Slovenia\\
\vspace{0cm}
$^3$Department of Physics, Indian Institute of Technology Madras, Chennai 600 036, India\\
\vspace{0cm}
$^4$Faculty of Civil and Geodetic Engineering, University of Ljubljana, 1000 Ljubljana, Slovenia\\
\vspace{0cm}
$^5$Institute of Mathematics, Physics and Mechanics, 1000 Ljubljana, Slovenia\\
$^6$ISIS facility, Rutherford Appleton Laboratory, Chilton, Didcot, OX11 0QX, Oxfordshire, UK\\
\vspace{0cm}
$^7$National High Magnetic Field Laboratory, Florida State University, Tallahassee, Florida 32310, USA\\
}

\end{center}
\end{widetext}
\subsection{1. Crystal structure refinement}

\begin{table}[h!]
\caption{{\bf Supplementary Table~I $\mid$ Crystal structure of \Nd.}
Atomic positions determined from Rietveld refinement of the powder X-ray diffraction pattern shown in Supplementary Fig.\,\ref{figS1}. 
The corresponding space group is $P\bar{6}c2$ (No.~188) and the cell parameters are $a = b = 6.2242(8)$\,\AA, $c = 19.938(3)$\,\AA, $\alpha =  \beta=90^\circ$, $\gamma = 120^\circ $.
\label{struct}}
\begin{ruledtabular}
\begin{tabular}{c|c|c|c|c|c}
 Atom & Wychoff & $x$  & $y$  & $z$  & Occ.  \\
\hline
Nd  &  $2c$ &  0.33333  &  0.66667 & 0  & 1\\
Ta1  &  $2e$ &  0.66667  & 0.33333  & 0  & 1\\
Ta2  &  $12l$ &  0.360(9)  &  0.358(6) & 0.1558(5)  & 1\\
O1  &  $12l$ &  0.21(6)  &  0.95(7) & 0.158(6)  & 1\\
O2  &  $12l$ &  0.40(5)  &  0.031 & 0.945(2)  & 1\\
O3  &  $6k$ &  0.41(2)  &  0.36(4) & 0.25  & 1\\
O4  & $4i$ &  0.66667  &  0.33333 & 0.17(2)  & 1\\
O5  & $4h$ &  0.33333  &  0.66667 & 0.13(5)  & 1\\
\end{tabular}
\end{ruledtabular}
\end{table}

\begin{figure}[h!]
\centering
\includegraphics[trim = 0mm 0mm 0mm 2mm, clip, width=1\linewidth]{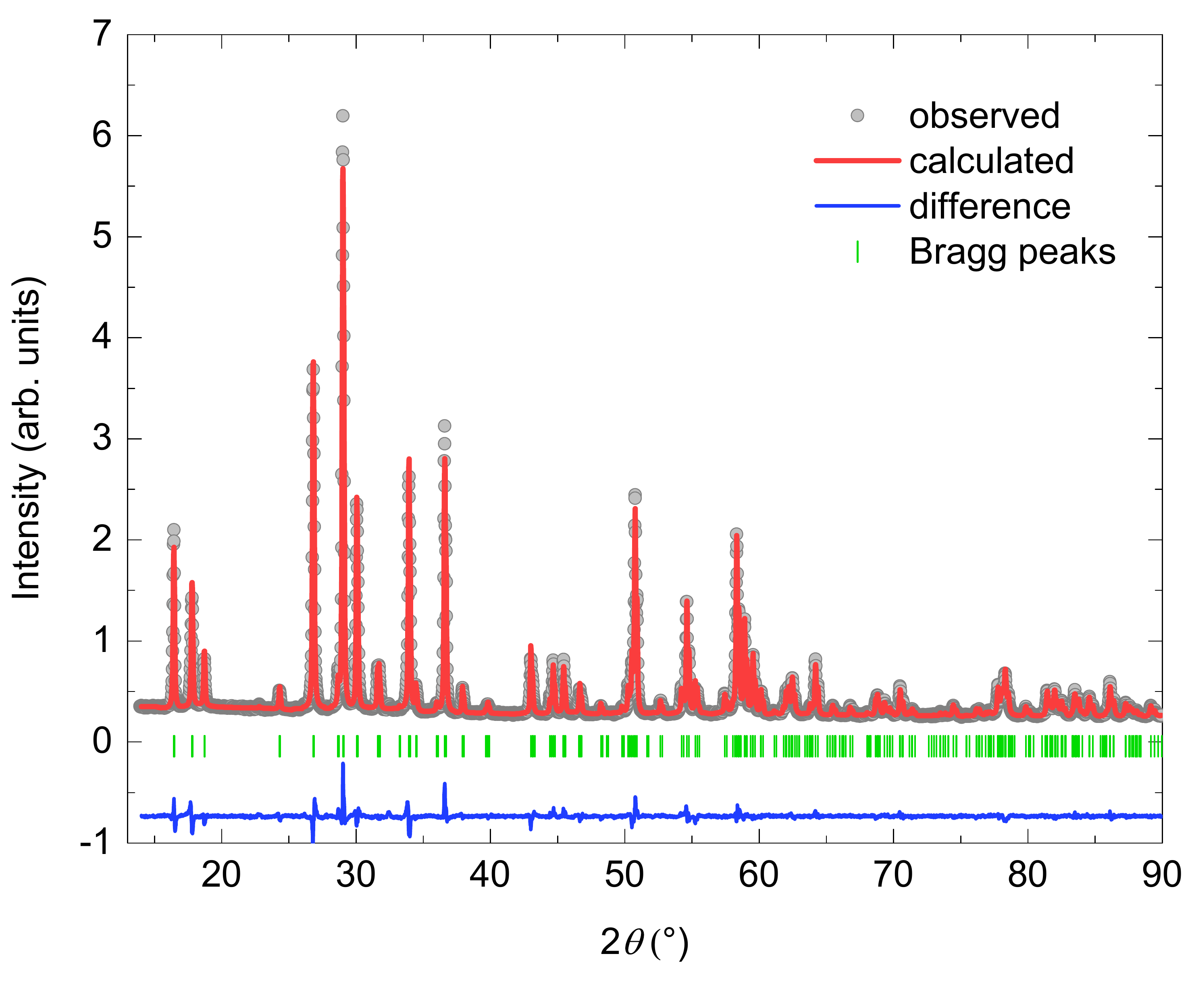}
\caption{{\bf Supplementary Fig.~1 $\mid$ Crystal structure.}
Rietveld refinement of powder XRD data of \Nd~within the space group $P\bar{6}c2$ (No.~188) at room temperature with $\chi^2 = 6.45$, $R_p = 10.9$, $R_{wp} = 10.2$ and $R_e = 4.04$.}
\label{figS1}
\end{figure}
 
The Rietveld refinement of the crystal structure of \Nd~based on its XRD pattern at room temperature is shown in Supplementary Fig.\,\ref{figS1}.
The refinement parameters summarized in Supplementary Table\,\ref{struct} are consistent with the previously reported values \cite{schaffrath1990chemischen}.
A structure with the Nd$^{3+}$ magnetic ions occupying a perfect triangular lattice (Wyckoff site $2c$) is found. 
All atomic sites are 100\% occupied and there is no intersite mixing.

\subsection{2. INS at incident energy of 100\,meV}
 
\begin{figure}[b]
\centering
\includegraphics[trim = 0mm 0mm 0mm 10mm, clip, width=1\linewidth]{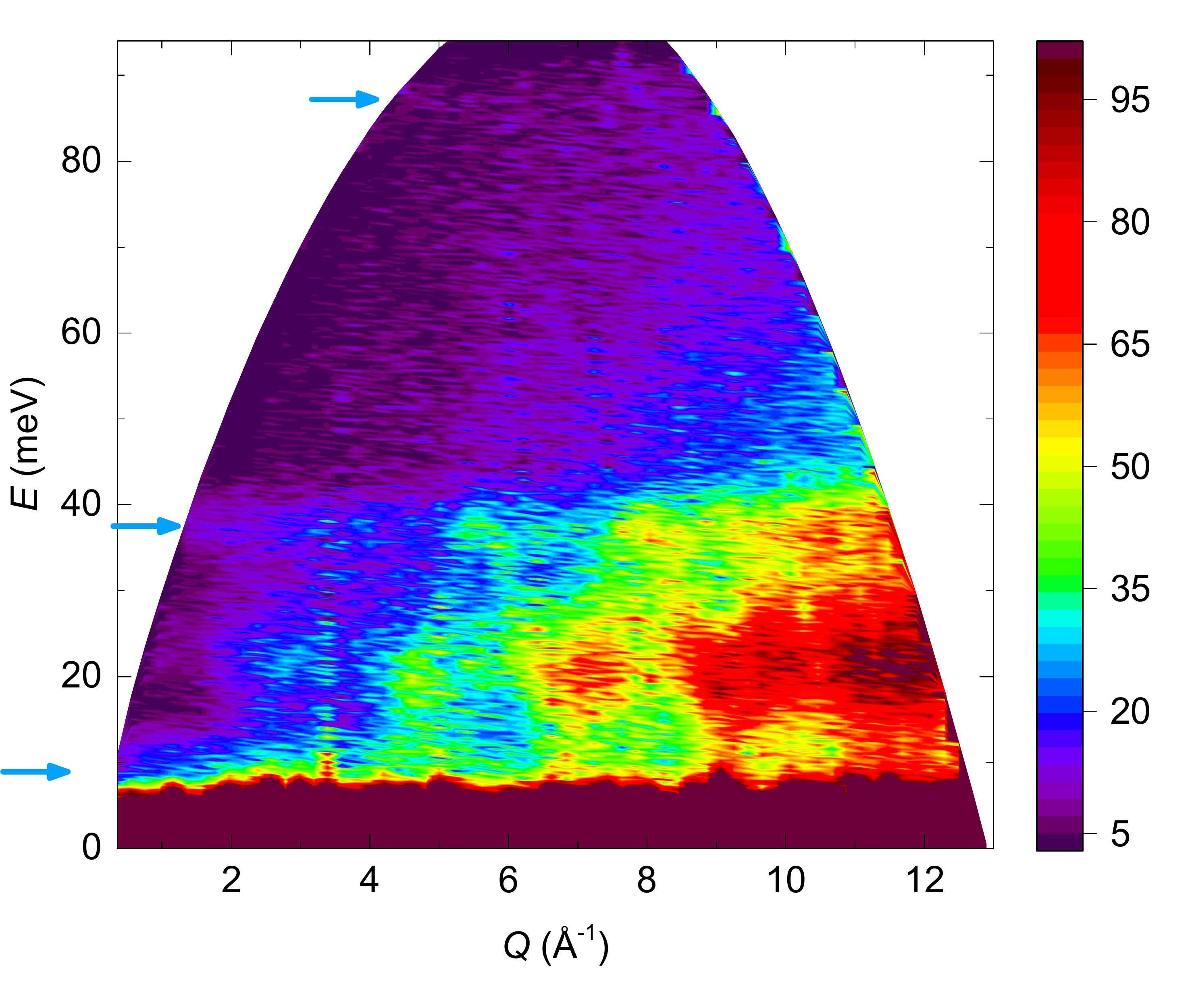}
\caption{{\bf Supplementary Fig.~2 $\mid$ Magnetic and phonon excitations in \Nd.}
INS intensity at 5\,K for neutron incident energy of 100\,meV. Arrows indicate the positions of the CEF transitions.}
\label{figS2}
\end{figure}
The dynamical structure factor measured at 5\,K by inelastic neutron scattering (INS) at the incident energy of $E=100$\,meV is shown in Supplementary Fig.\,\ref{figS2}.
Contrary to the measurements shown in Fig.\,3a in the main document, which is limited to $E=60$\,meV, an extra crystal-electric-field (CEF) excitation is observed at 87~meV at the smallest scattering wave vectors $Q$.
This transition is much weaker than the other two CEF transitions observed at lower energies and smaller $Q$'s (indicated by arrows in Supplementary Fig.\,\ref{figS2}), because of the magnetic form factor of Nd$^{3+}$, which significantly reduces with increasing Q (see inset in Fig.\,3b in the main document). 
The dominant scattering above the quasi-elastic line (limited to 7\,meV) comes from phonons.

\begin{figure}[t]
\centering
\includegraphics[trim = 0mm 0mm 0mm 0mm, clip,width=\linewidth]{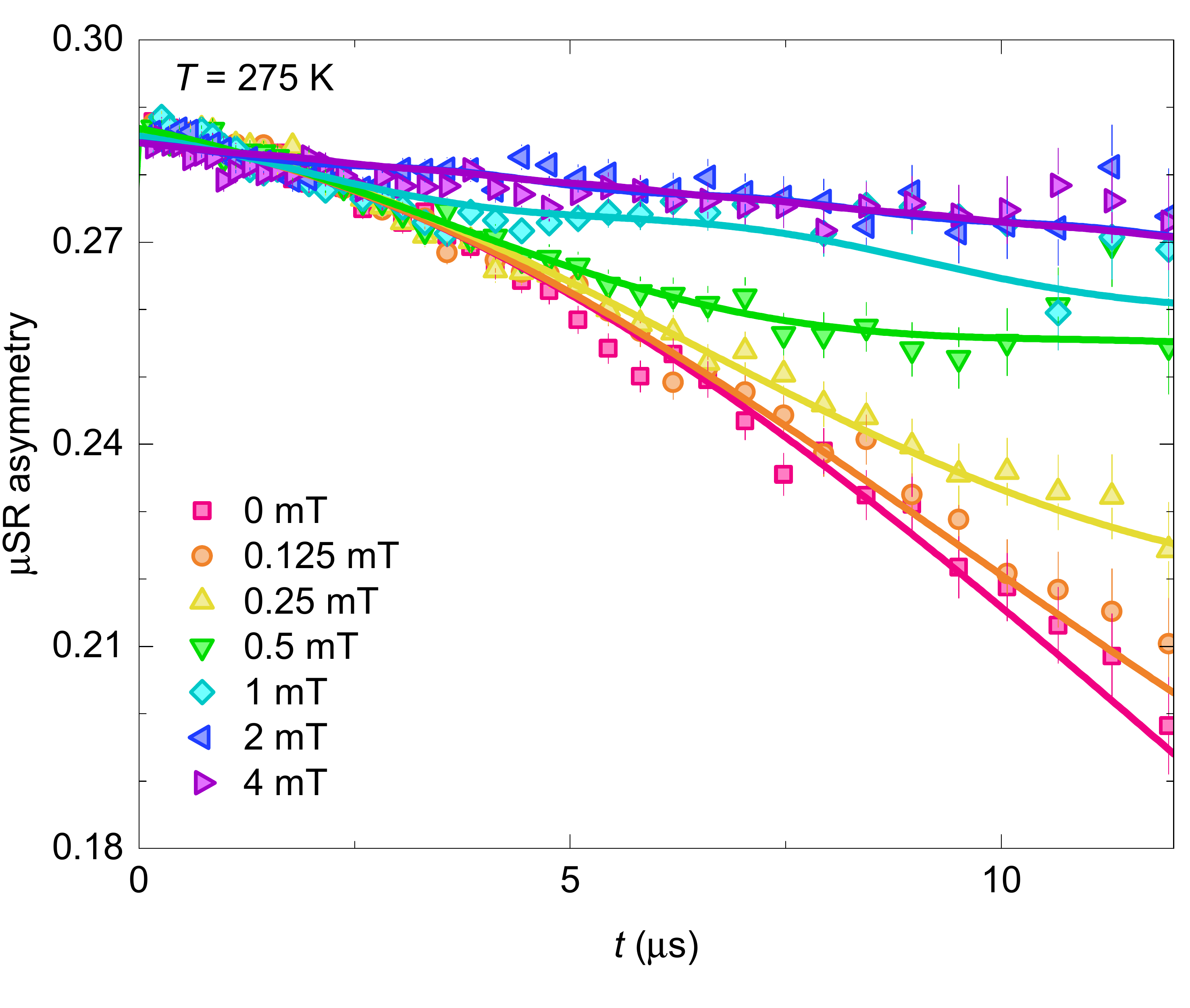}
\caption{{\bf Supplementary Fig.~3 $\mid$ $\mu$SR field-decoupling experiment.}
Muon spin asymmetry measured in various longitudinal fields at 275\,K. The solid lines correspond to equation (\ref{FD}), with the muon polarization $P(t)$ given by equation (6) in the main text.}
\label{figS3}
\end{figure}

\subsection{3. $\mu$SR at 275\,K}
A $\mu$SR field-decoupling experiment was performed at 275\,K to separate the relaxation of the $\mu$SR asymmetry $A(t)$ caused by static nuclear fields from relaxation due to dynamical local field of electronic origin at the muon stopping site in \Nd, and to determine the background signal arising from muons stopping outside the sample.
The data collected in different longitudinal applied fields spanning the interval from 0 to 4\,mT was simultaneously fit with the model 
\begin{equation}
A(t) = A_0 P(t) + A_{\rm bgd},
\label{FD}
\end{equation}
where the muon polarization $P(t)$ follows equation\,(6) in the main document.
The fit parameters corresponding to fits (solid lines) in Supplementary Fig.\,\ref{figS3} are summarized in the Methods section of the main document.
It is found experimentally that the applied field already of the order of 1\,mT completely removes the relaxation due to static nuclear field, and the remaining relaxation of the $\mu$SR asymmetry is entirely due to dynamical local field.
This relaxation is not affected by the applied field.

\end{document}